# Electroluminescence from isolated defects in zinc oxide – towards electrically triggered single photon sources at room temperature

Sumin Choi, Amanuel M. Berhane, Angus Gentle, Cuong Ton-That, Matthew R Phillips, and Igor Aharonovich*

*School of Physics and Advanced Materials, University of Technology Sydney, 15 Broadway, Ultimo, New South Wales 2007, Australia*

Single photon sources are required for a wide range of applications in quantum information science, quantum cryptography and quantum communications. However, so far majority of room temperature emitters are only excited optically, which limits their proper integration into scalable devices. In this work, we overcome this limitation and present room temperature electrically triggered light emission from localized defects in zinc oxide (ZnO) nanoparticles and thin films. The devices emit at the red spectral range and show excellent rectifying behavior. The emission is stable over an extensive period of time, providing an important prerequisite for practical devices. Our results open up possibilities to build new ZnO based quantum integrated devices that incorporate solid-state single photon sources for quantum information technologies.

Email: igor.aharonovich@uts.edu.au

Single-photon sources (SPSs) that generate non-classical states of light have been extensively explored over the past decade due to a variety of applications including quantum cryptography, quantum computation, spectroscopy and metrology[1-3]. While sources based on quantum dots (QDs) are robust and have ideal optical properties, such as narrow emission line-width and short excited state lifetime, their operation is mostly limited to cryogenic temperatures[4-6]. Alternatively, color centers in wide band gap materials are excellent candidates for room temperature SPSs. These color centers are defects within the band gap of the host matrix, and form localized confined states that can emit single photons on demand. Diamond, for example, has been studied extensively, due to its ability to host plethora of emitters that are photostable and exhibit single photon emission at room temperature[7].

The need to integrate the SPSs with scalable photonic devices, such as resonators or optical cavities, stems the urgent need for realization of electrically triggered SPSs [8-12]. It has been shown previously that electrically driven light emission was realized with a heterojunction nanostructures, where carrier injection occurs across the p-n junction or in the i region of a p-i-n junction[13-15]. However, engineering efficient junctions from materials like diamond is challenging, and requires sophisticated growth conditions and cumbersome implantation of the single emitting defects[16].

A more promising approach for fabrication of quantum light emitting diodes (LEDs) is exploiting other semiconductors that are more suitable for current optoelectronic applications. One of these materials is zinc oxide (ZnO), which has recently been shown to host single defects that are harnessed as SPSs[17-19]. In addition, ZnO has attracted significant attention for its extensive photonic applications in ultraviolet and visible spectral range due to its relatively large bandgap (3.37 eV) and high exciton binding energy (60 meV)[20]. The mature technology of ZnO heterojunctions with silicon or gallium nitride has enabled fabrication of advanced optoelectronic devices including transistors and LEDs[21-24]. Therefore, the transformation of these technologies into the quantum regime, where single emitters can be electrically addressed in ZnO heterojunctions is a promising avenue for scalable quantum photonic applications.

In this work we report efficient electrically driven light emission from individual defects in a n-ZnO/p-Si heterojunctions. We investigate two different sources of n-ZnO: the first source is based on sputtered ZnO films and the second involves deposition of ZnO nanoparticles. n-ZnO/p-Si heterojunction devices have been chosen due to their cost effective and mature fabrication techniques.

A schematic diagram of the devices is shown in Figure 1a. First, ~2 mm diameter circular electrodes were patterned by standard lithographic process. Then, 300 nm SiO$_2$ layer was deposited on a p-Si substrate (Boron-doped, 0.001-0.005 ohm·cm) via e-beam evaporation, followed by 150 nm Al sputtered as the electrodes on the SiO$_2$ layer to achieve a Shotky barrier. The thin layer of SiO$_2$ is used as a spacer between the two electrodes. The photoresist was lifted-off by ultrasonication in acetone solution for 5 minutes to fabricate the proper structure of Si/SiO$_2$/Al.

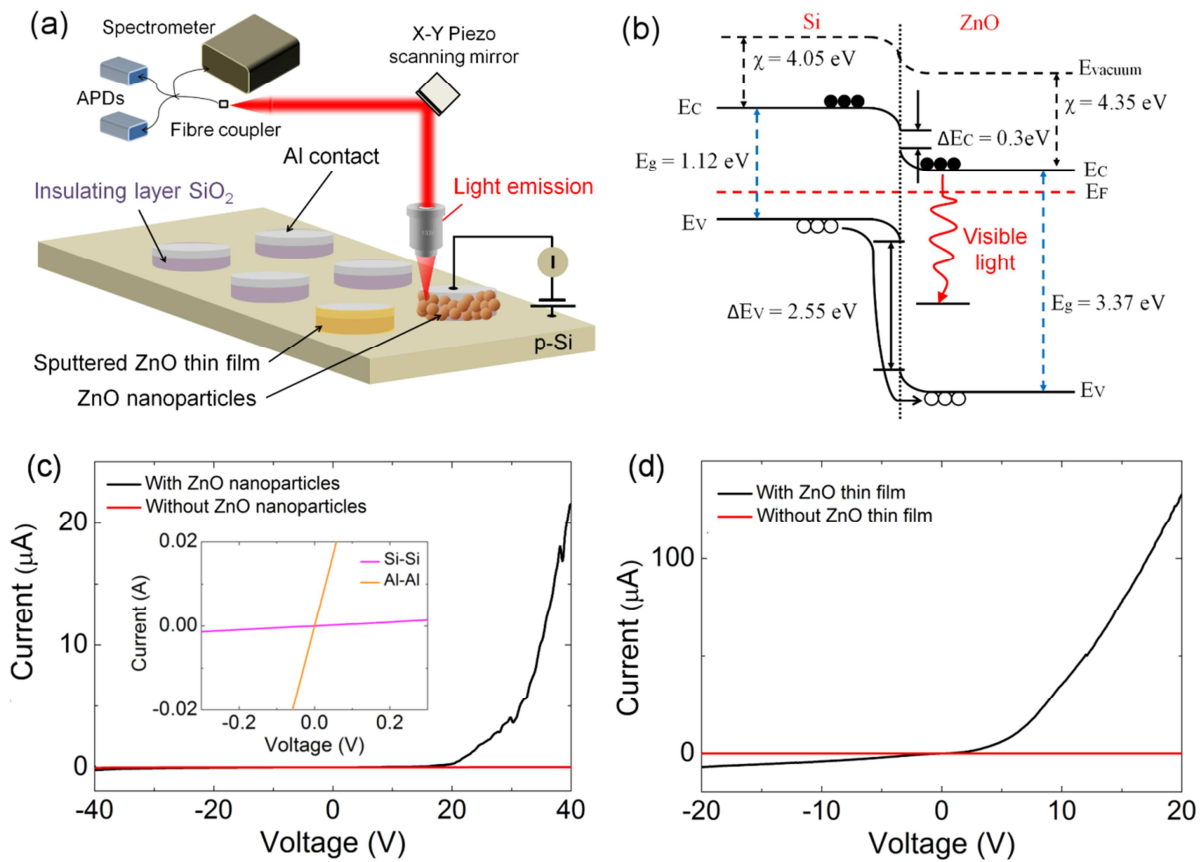

*Figure 1. (a) Schematic diagram of the n-ZnO/p-Si heterojunction. Electrically driven light emission is generated at the edge of the circle, and then collected through a microscope objective and directed into APDs or a spectrometer. (b) The energy band diagram of the n-ZnO/p-Si heterojunction under zero voltage bias. Defect-related radiative recombination occurs in the devices. The electrons in the conductions band of ZnO will drop down into the defect-related energy level of ZnO to recombine with the holes therein, giving rise to the visible emission. (c) I-V characteristic of n-ZnO/p-Si heterostructure devices; (a) ZnO nanoparticles/Si configuration with a threshold voltage of ~ 18 V and a measured current of 20 μA at 40 V of forward bias. Inset is the I-V characteristics of Al-Al and Si-Si contacts showing good ohmic characteristics. (d) ZnO thin film/Si configuration, with a threshold voltage of ~7 V and a measured current of 200 μA at 20 V for the ZnO thin film-based device.*

To fabricate the heterojunction with the nanoparticles, ZnO nanoparticles (20 nm, Nanostructured and Amorphous Materials Inc., USA) were first annealed at 500 °C for 30 minutes, dispersed in methanol solution and then drop cast on to the pre-patterned Si/SiO$_2$/Al substrate to stick on the wall of the mesa to obtain p-Si/n-ZnO heterojunction. To engineer the devices with sputtered ZnO, 50 nm ZnO thin films were grown on the wall of the mesas using a 0.25A DC current controlled deposition from a 2" ZnO (99.99%) sputter target with an Argon pressure of 2 mTorr. Prior to the deposition, the sputter chamber was evacuated to a base pressure of ~ $5\times 10^{-7}$ torr. The film thickness was monitored using a quartz crystal monitor in the chamber. To obtain reliable Ohmic contacts and to create the color centers within the sputtered ZnO, the sample was annealed at 500 °C for 30 minutes in air environment.

To understand the mechanism of light generation from the formed devices, an energy band alignment diagram n-ZnO/p-Si heterojunction is considered in Figure 1b based on individual band structures. The dominant mechanism of EL is the recombination of holes injected from the Si with electrons in the ZnO, supplied additionally by the Al contact. With the increased forward bias, the energy barriers for electrons and holes are both lowered, thus favoring the injection of electrons and holes. The electron affinities of Si ($\chi_{Si}$) and ZnO ($\chi_{ZnO}$) are 4.05 eV and 4.35 eV, respectively while the bandgap energies are 1.12 eV and 3.37 eV, for Si and ZnO, respectively. Therefore, the conduction band offset for electrons is $\Delta E_C = \chi_{ZnO} - \chi_{Si} = 0.3$ eV, while that for holes is $\Delta E_V = 2.55$ eV[25]. Although holes injected from the p-Si are limited due to the large barrier, the very high concentration of holes in p-Si causes a certain amount of holes to be injected into ZnO under the appropriately high forward bias. The electrons from the conduction band of the ZnO may first occupy these empty defect-produced traps and subsequently, directly recombine with deep level defects in the band-gap to produce the visible emissions. The details of the mechanism are discussed in later with EL and PL spectra from the devices.

To study the diode characteristics, current – voltage (I-V) measurements were carried out for the two different devices and are shown in Figure 1c-d. Tungsten probes with 1 µm tips attached to micro-positioners were used to connect to the sample electrodes. The inset of Figure 1c shows a linear behavior of I-V characteristic between two Si-Si and Al–Al pots, indicating that a good Ohmic contact was achieved. The red curves represent the measurement without the ZnO materials to confirm that there is no metal leakage of Al on the Si through the insulating layer of SiO$_2$. The black curves in figure 1c (d) are the I-V measurements after the deposition of ZnO nanoparticles (film) on the samples. Both junctions exhibit excellent, well-defined rectifying behavior. The ZnO nanoparticle device shows a threshold voltage at ~18 V and a forward current more than 20 μA at 40V, whereas the ZnO thin film device exhibits higher current, ~200 μA at 20 V with lower threshold voltage of ~7 V. The different onset voltages between two devices and noise of the curves in Figure 1c may arise from the presence of surface states and/or presence of an oxide layer at the interface as the SiO$_2$ layer that acts as a barrier in series[26] as well as unavoidable voids in the nanoparticle sample.

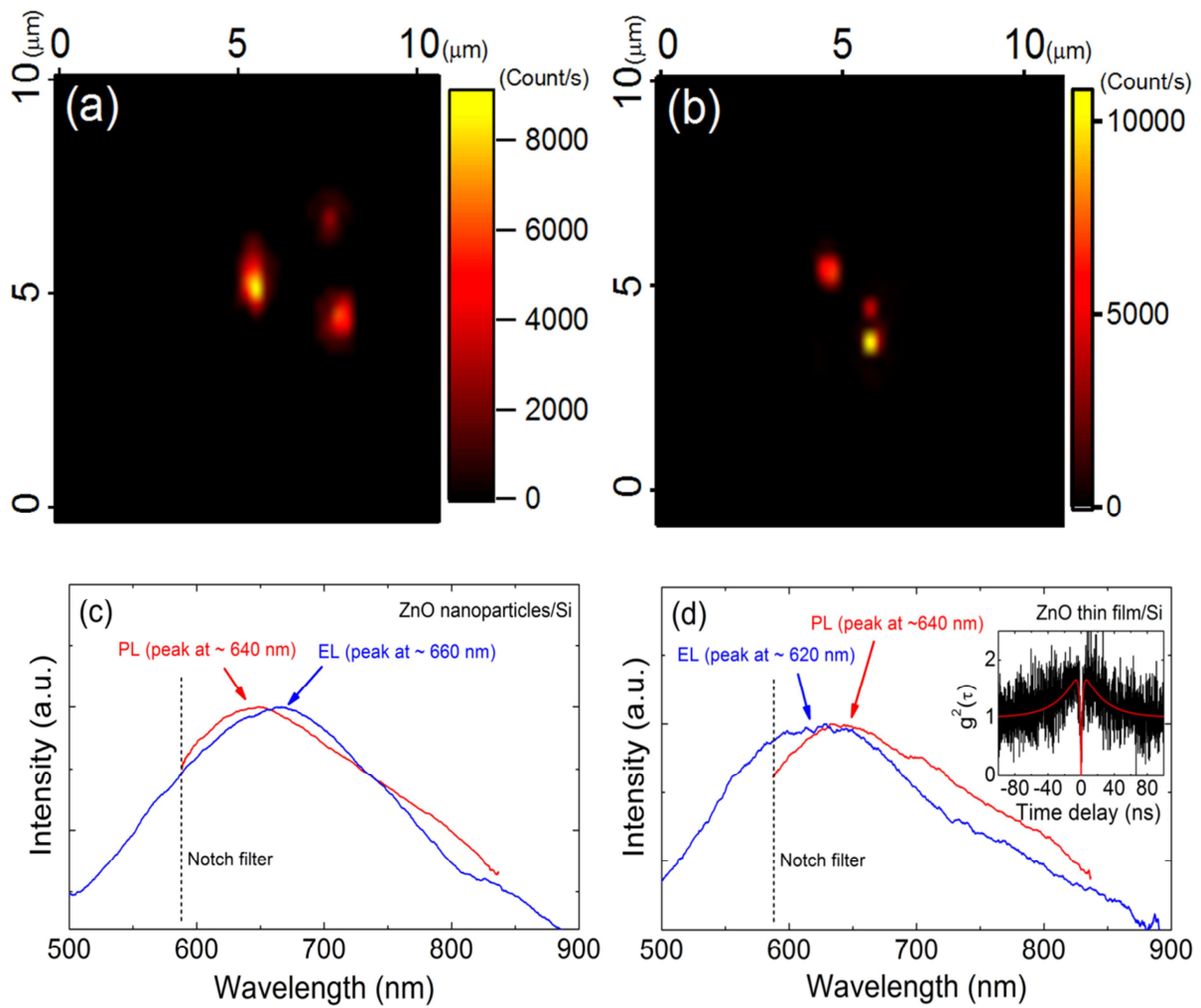

*Figure 2. EL and PL of defects in ZnO/Si devices recorded at room temperature. (a) EL confocal maps recorded from the ZnO nanoparticles/Si and (b) ZnO film devices, respectively. The bright spots correspond to defect-related color centers in ZnO. (c,d) EL and PL spectra of the devices recorded at room temperature; (c) ZnO nanoparticles/Si, and (d) ZnO thin film/Si devices. Both devices exhibit orange-red emission ranging from ~550 nm to 800 nm when 40 V and 15 V were applied to ZnO nanoparticles/Si and ZnO thin film/Si, respectively. While PL spectra show no difference from both samples, peak wavelength of EL are slightly different, possibly resulting from different defect centers in ZnO. Inset of Figure 3 (d) is the*

*second-order autocorrelation function $g^2(\tau)$ of ZnO thin film excited by PL, indicating the presence of a single quantum emitter in the ZnO thin film. The bunching ($g^2(\tau) > 1$) indicates of a presence of a metastable state.*

To investigate the luminescent properties of the formed devices, electroluminescence (EL) and photoluminescence (PL) measurements were collected using a confocal microscope with 500 nm lateral resolution. The signal was collected through an objective with a numerical aperture of 0.7 and directed into a spectrometer (Princeton Instruments, 300 lines/nm grating). For the PL excitation, a continuous wave laser of 532 nm was employed. A dichroic mirror was used to filter the excitation laser. All signals were recorded using an Avalanche Photo Diode (Excelitas, SPCM-AQRH-14) and analyzed using a Single Photon Counting (PicoHarp 300). The measurements were carried out at room temperature under ambient conditions. Upon applying the voltage between the Al contacts and the Si wafer, the emission was generated at the edge of the circular mesas. Throughout the measurements, the contact probes were positioned close to the scanning area to achieve higher EL generation and therefore better signal to noise.

Figure 2 (a,b) shows room temperature EL confocal maps recorded from the ZnO nanoparticles and the sputtered ZnO heterojunctions, respectively. The bright spots in the confocal image correspond to individual electrically excited luminescence defects. For the ZnO nanoparticles-based device, the EL signal becomes detectable when a forward direct current bias of 35 V is applied across the device. No EL is detected under reverse biasing. However, further increase of applied voltage over 40 V resulted in electrical breakdown of the devices.

Figure 2 (c,d) show the corresponding EL spectra from the nanoparticles and the sputtered films (blue curves). Both devices exhibit broad peaks at the red spectral range (at 660 nm and 620 nm) – that is typical to the sub bandgap defect emission[18]. The fact that EL signal was obtained from a localized spot indicates that recombination occurs on an individual defect site – ideal for generation of non-classical light. Neither sample exhibited exciton-related near band edge emissions in the ultraviolet region, possibly due to the self-absorption by Si substrate or deep-level traps in ZnO.

Complementary PL spectra from same spots were collected using a 532 nm excitation (with no bias applied to the sample) to access the deep level defects in the ZnO. The spectra are shown in Figure 2 (c,d) as red curves. Slight spectral shift between the EL and the PL signals can be observed, likely due to different excitation pathways. However, their full width half maximum spectra are similar, indicating that similar defects are addressed. The EL and PL signals can be potentially ascribed to oxygen interstitials centers or single ionized zinc vacancy defects[27, 28].

To verify that the ZnO thin film exhibits single photon emission, Hanbury-Brown Twiss (HBT) interferometer was used on the same position of the confocal map in Figure 2b under a 532 nm laser excitation. Inset of figure 2 (d) shows the second order correlation function $g^2(\tau)$ from the ZnO defect center. An antibunching dip at zero delay time ($g^2(\tau) < 0.5$) indicates that the emission originates from a single photon emitter. Because the emitter was measured at high excitation power (~ 2 mW), bunching behavior was also observed, indicative of a three-level system with a shelving (metastable) state. The red line is the theoretical fit using a three level system equation, $g^2(\tau) = 1 - (1 + a) exp(-\lambda_1 \tau) + a\, exp(-\lambda_2 \tau)$, where $\lambda_1$ and $\lambda_2$ are decay rates for radiative and the metastable states, respectively.

We then studied the emission saturation from the electrically driven devices. Figure 3 (a, b) shows the measured count rate as a function of injection current for the ZnO nanoparticles/Si and the ZnO thin film/Si, respectively. The black dots are the experimental values, while the red curves are the fit according to the following equation, $C = C_{Sat} I/(I_{Sat} + I)$, where C is the emission count rate at a given injection current ($I$) and $C_{Sat}$ represents saturation count rates at saturation current ($I_{Sat}$). From the fitting equation, the saturation count rates are determined to be $7.23 \times 10^3$ count/s and $2.07 \times 10^4$ count/s with saturation current of 32.4 μA and 6.4 μA for the ZnO nanoparticles/Si and ZnO thin film/Si, respectively. These values are comparable with other electrically driven sources[8, 9].

Finally, to test the photostability of the studied defects, luminescence intensity traces were measured as a function of applied current. Figure 3(c, d) shows the stability measurements of the EL generated from the ZnO nanoparticles and the ZnO film devices, respectively. These figures show that the emission was persistently stable more than 30 minutes, indicating that the ZnO/Si devices could be a potential light source for future solid-state quantum photonic applications operating at room temperature.

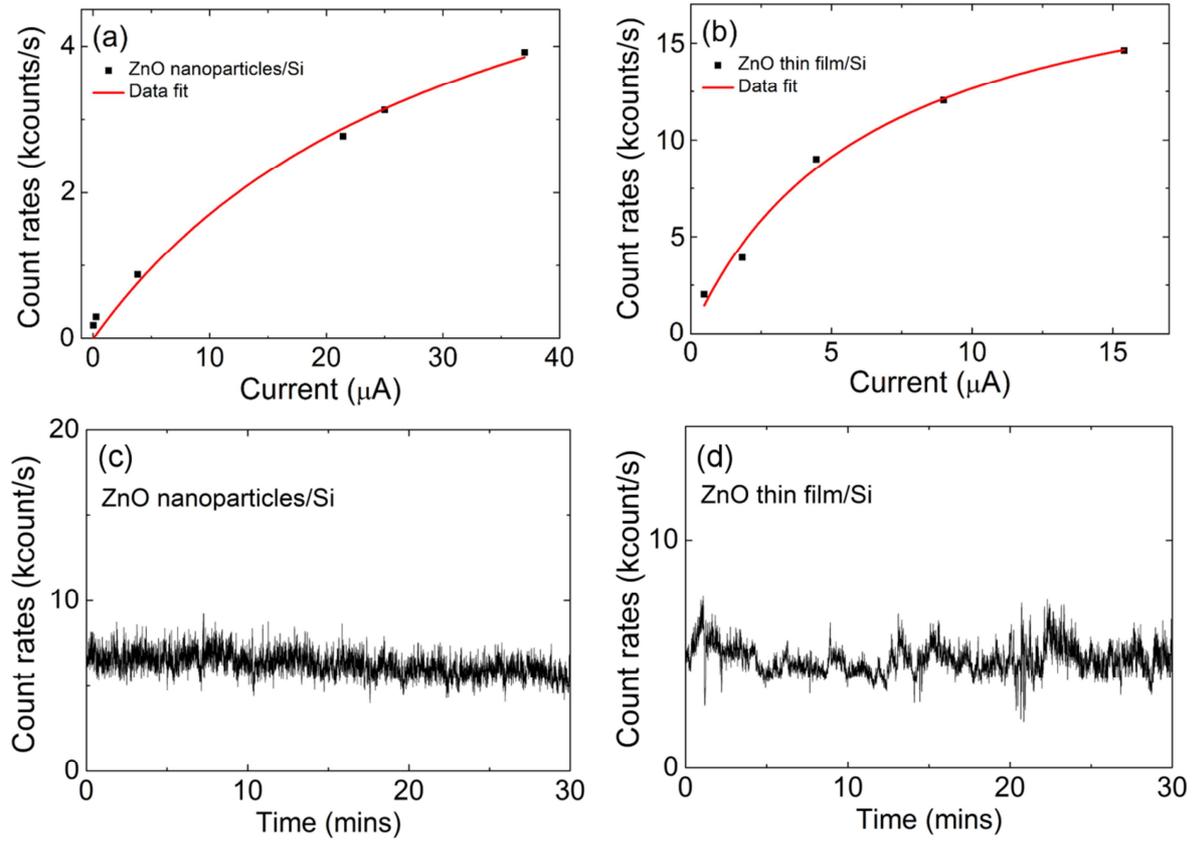

*Figure 3. (a) Count rate of the EL generated emission as a function of the device current for the ZnO nanoparticles/Si and (b) ZnO thin film/Si devices. The black dots are raw data and red curves are the fitting curve showing saturation behaviors according to the Equation (1) with $C_{Sat} = 7.2 \times 10^3$ count/s $I_{Sat} = 32.4$ µA for ZnO nanoparticles/Si and $C_{Sat} = 2.1 \times 10^4$ count/s $I_{Sat} = 6.4$ µA for ZnO thin film/Si, respectively. (c) and (d) are the intensity traces recorded from one of the bright spots in the confocal map from the ZnO nanoparticcles/Si and the ZnO thin film/Si devices, respectively . Both devices exhibited excellent photostability for more than 30 minutes.*

In summary, we reported on electrically modulated emission from isolated ZnO defects, integrated within a ZnO/Si heterojunction. The devices were fabricated either from ZnO nanoparticles or from sputtered ZnO. Room temperature I-V characteristics of the diodes confirmed excellent rectifying behavior with the threshold voltages at ~18 V and ~7 V for ZnO nanoparticles and thin film devices, respectively. Defect-related electroluminescence at the red spectral range has been achieved under forward bias, with moderate count rate of several kcounts/s at saturation. In addition, both devices were stable over 30 minutes under the high forward bias, which is crucial for the development of future ZnO based quantum devices. In combination with the recent progress into ZnO cavities and resonators[29, 30], our results pave the way to scalable and cost efficient fabrication of electrically driven, quantum nanophotonic devices employing ZnO as the fundamental building block.

ACKNOWLEDGEMENTS

Igor Aharonovich is the recipient of an Australian Research Council Discovery Early Career Research Award (Project No. DE130100592). The authors would like to thank G. McCredie for technical support and Olga Shimoni for helpful discussions.

**Notes**

The authors declare no competing financial interest.